\documentclass[aps,prb,twocolumn,superscriptaddress]{revtex4-1}
\usepackage{amsmath,amsthm,amssymb}
\usepackage[dvipdfmx]{graphicx}
\usepackage{amsmath,bm}
\usepackage{color,ulem}

\newcommand{\1}{\mbox{1}\hspace{-0.25em}\mbox{l}}

\newlength{\figwidth}
\newlength{\figlarge}
\begin{document}
%%%%%%%%%%%%%%%%%%%%%%%%%%%%%%%%%%%%%%%%%%%%%%%%%%%%%%%%%%%%%%%%%%%%%%%
\title{
Topological edge Mott insulating state in two dimensions at finite temperatures\\
-bulk and edge analysis-
}
%%%%%%%%%%%%%%%%%%%%%%%%%%%%%%%%%%%%%%%%%%%%%%%%%%%%%%%%%%%%%%%%%%%%%%%
\author{Tsuneya Yoshida}
\affiliation{Condensed Matter Theory Laboratory, RIKEN, Wako, Saitama, 351-0198, Japan}
\author{Norio Kawakami}
\affiliation{Department of Physics, Kyoto University, Kyoto 606-8502, Japan}

%%%%%%%%%%%%%%%%%%%%
%%%%%%%%%%%%%%%%%%%%%%%%%%%%%%%%%%%%%%%%%%%%%%%%%%%%%%%%%%%%%%%%%%%%%%%
\date{\today}
%%%%%%%%%%%%%%%%%%%%%%%%%%%%%%%%%%%%%%%%%%%%%%%%%%%%%%%%%%%%%%%%%%%%%%%
\begin{abstract}
We study a bilayer Kane-Mele-Hubbard model with lattice distortion and inter-layer spin exchange interaction under cylinder geometry.
Our analysis based on real-space dynamical mean field theory with continuous-time quantum Monte Carlo demonstrates the emergence of a topological edge Mott insulating (TEMI) state which hosts gapless edge modes only in collective spin excitations. This is confirmed by the numerical calculations at finite temperatures for the spin-Hall conductivity and the single-particle excitation spectrum; the spin Hall conductivity is almost quantized, $\sigma^{xy}_\mathrm{spin}\sim2(e/2\pi)$, predicting gapless edge modes carrying the spin current, while the  helical edge modes in the single-particle spectrum are gapped out with respecting symmetry. 
It is clarified how the TEMI state evolves from the ordinary spin Hall insulating state with increasing the Hubbard interaction at a given temperature and then undergoes a phase transition to a trivial Mott insulating state. With a bosonization approach at zero temperature, we further address which collective modes host gapless edge modes in the TEMI state.
\end{abstract}
%%%%%%%%%%%%%%%%%%%%%%%%%%%%%%%%%%%%%%%%%%%%%%%%%%%%%%%%%%%%%%%%%%%%%%%
%%%%%%%%%%%%%%%%%%%%%%%%%%%%%%%%%%%%%%%%%%%%%%%%%%%%%%%%%%%%%%%%%%%%%%%

%%%%%%%%%%%%%%%%%%%%%%%%%%%%%%%%%%%%%%%%%%%%%%%%%%%%%%%%%%%%%%%%%%%%%%%
\maketitle
%%%%%%%%%%%%%%%%%%%%%%%%%%%%%%%%%%%%%%%%%%%%%%%%%%%%%%%%%%%%%%%%%%%%%%%

%%%%%%%%%%%%%%%%%%%%%%%%%
\section{Introduction} 
%%%%%%%%%%%%%%%%%%%%%%%%%

Topological insulators in free fermion systems host gapless edge/surface states \cite{TI_review_Hasan10,TI_review_Qi10}, which can be a source of various remarkable properties. For instance, three-dimensional strong topological insulators with time-reversal symmetry show topological magnetoelectric effects due to gapless Dirac cones at the surface\cite{Qi_3DTI2008}.
At the ends of one-dimensional topological superconducting chains without time-reversal symmetry, Majorana fermions emerge, whose experimental confirmation is a subject of intensive studies\cite{Mourik_Majorana2012,Majorana_Rokhinson2012,Majorana_Anindya2012}.

One of the important issues of topological insulators is the impact of electron correlations. Recently, first principle calculations suggest that $4f$- ($5d$-) electron compounds, such as $\mathrm{SmB_6}$\cite{Takimoto_SmB6_2011}, $\mathrm{Pr_2Ir_2O_7}$\cite{TMI_LBalents09,Yang_R2Ir2O7_STI_2010,Kargarian_R2IrO7_STI_2011} \textit{etc.}, can be strong topological insulators in correlated systems.
In these systems, electron correlations and topological properties are expected to trigger off novel phenomena, and this issue is extensively studied both theoretically and experimentally.

As the results of these extensive studies, it is elucidated that interaction effects can dramatically change the properties of gapless edge modes\cite{Z_to_Zn_Fidkowski_10,YaoRyu_Z_to_Z8_2013,TMI_LBalents09}.
In particular, Pesin \textit{et al.}\cite{TMI_LBalents09} suggests the emergence of a novel topological state in three-dimensional topological insulators with electron correlations.
According to their analysis based on a slave-boson mean field theory, this new phase hosts gapless modes not in the single-particle spectrum but in the spin excitation spectrum (i.e., a collective excitation spectrum) due to the interplay of electron correlations and topologically nontrivial properties in the bulk. 
In this paper, we refer to this topological state as a topological edge Mott insulating (TEMI) state.
The TEMI state arising from electron correlations attracts much interest, and several attempts have been made to address this issue beyond the slave-boson approach \cite{TopoMott_Yamaji2011,TBI_Mott_Tada,TopoMott_Yoshida2014,You_bosonization_2015}.
Unfortunately, so far, the realization of this exotic phase in two (three) dimensions is still under debate. 
Especially, there are few systematic analyses of bulk and edge properties supporting the realization of the TEMI state in a lattice model, although its possibility is discussed with an effective field theory for one-dimensional edges\cite{You_bosonization_2015}.

In order to address these issues, in this paper, we analyze a bilayer Kane-Mele-Hubbard model\cite{bilayerKane_Slagle_2015,bilayerKane_He_2015} with lattice distortion and spin exchange interaction. In the non-interacting case, the topological phase shows the quantized spin-Hall conductivity $\sigma^{xy}_\mathrm{spin}\sim2(e/2\pi)$, implying that electrons with up- and down-spin state propagate in opposite direction (helical modes) in each layer.
Our study reveals that in the presence of electron correlations, the helical modes in the single-particle spectrum are gapped, while the spin-Chern number remains unchanged in the bulk. These results demonstrate the possible realization of the TEMI state where the edge modes remain gapless only in the collective channels.  We also elucidate which collective channels host gapless edge modes by employing the bosonization approach.

The rest of this paper is organized as follows. 
In the next section (Sec.~\ref{sec: model_and_method}), we explain our model and method.  The obtained results are presented in Sec.~\ref{sec: results}, where we clarify how the TEMI state emerges in the correlated topological insulators, by using both of finite-temprerature and zero-temperature approaches.
A short summary is given in Sec. IV.

%%%%%%%%%%%%%%%%%%%%%%%%%
\section{Model and method}\label{sec: model_and_method}
\begin{figure}[!h]
\begin{center}
\includegraphics[width=\hsize,clip]{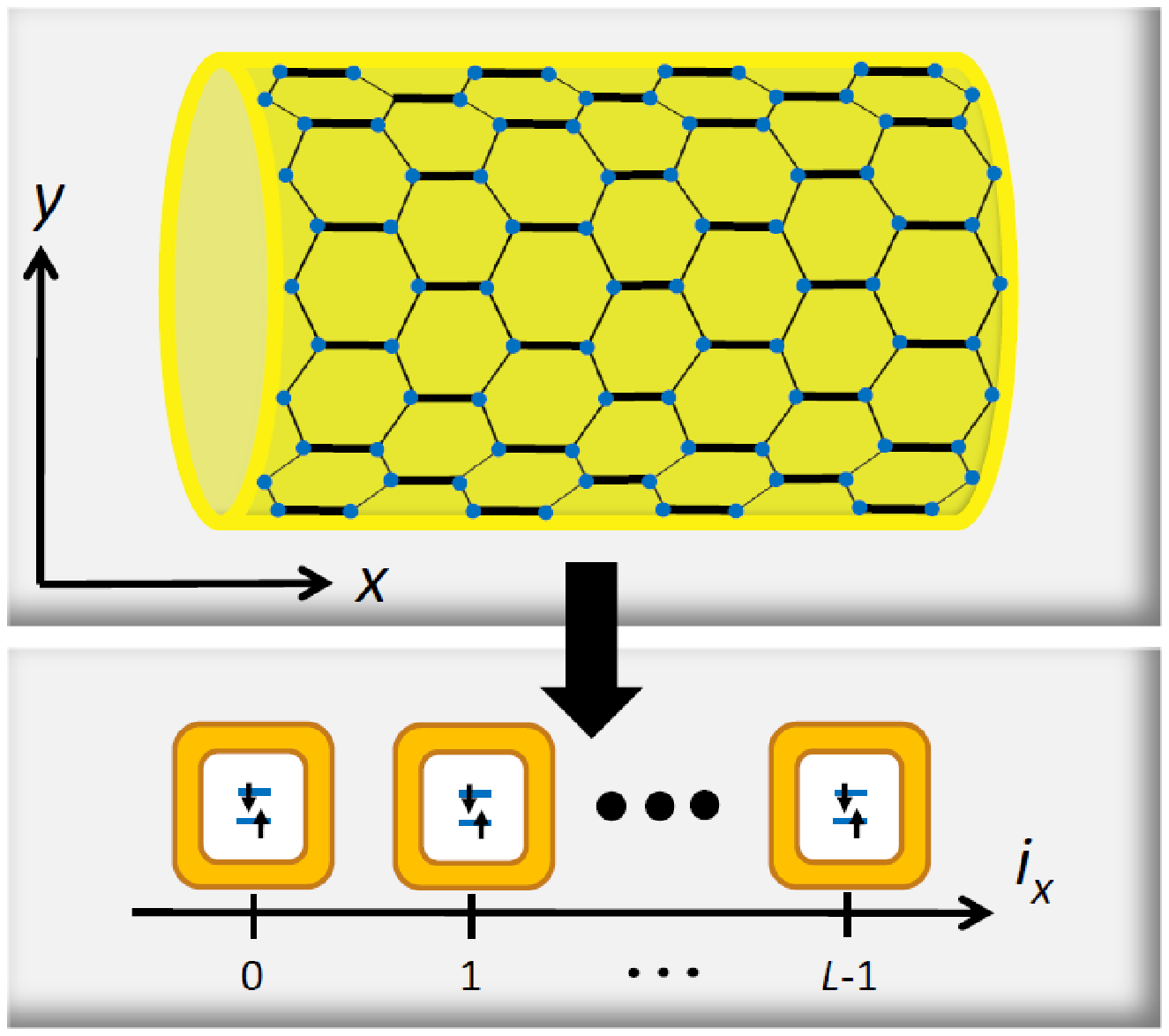}
\end{center}
\caption{(Color Online). 
Sketch of the hopping $t_{i,j}$. The thick (thin) lines denote the hopping $t$ ($rt$).
The lattice model is mapped to $L$-effective impurity problems, which yield the site dependent self-energy.
}
\label{fig:model}
\end{figure}

%%%%%%%%%%%%%%%%%%%%%%%%%
We study a bilayer Kane-Mele Hubbard model with lattice distortion and spin exchange interaction\cite{bilayerKane_Slagle_2015,bilayerKane_He_2015,bilayergrahane_Bi_2016}. 
The Hamiltonian reads
%%%%%%%%%%%%%%%%%%
\begin{subequations}
\begin{eqnarray}
H&=&H_0+H_\mathrm{int}, \\
H_0&=&-\sum_{\langle i,j\rangle,\alpha}t_{i,j}c^\dagger_{i,\alpha,\sigma}c_{j,\beta,\sigma} \\
   &&\quad+it_{so}\sum_{\langle\langle i,j\rangle \rangle }  \bm{\sigma}_{\sigma\sigma'} \cdot \bm{d}_i\times \bm{d}_j/|\bm{d}_i\times \bm{d}_j| c^\dagger_{i,\alpha,\sigma}c_{j,\beta,\sigma'}, \nonumber  \\
H_\mathrm{int}&=&U\sum_{i\alpha} n_{i,\alpha,\uparrow}n_{i,\alpha,\downarrow} +J\sum_{i}\bm{S}_a\cdot\bm{S}_b,
\end{eqnarray}
\end{subequations}
%%%%%%%%%%%%%%%%%%
where $\bm{d}_i$ and $\bm{d}_j$ are vectors connecting site $i$ and $j$.
$c^\dagger_{i,\alpha,\sigma}$ creates an electron with $\sigma=\uparrow,\downarrow$ state at site $i$ in layer $\alpha=a,b$.
$t_{i,j}$ is a hopping integral between neighboring sites $i$ and $j$. If the electron hops in $x$-direction, we set $t_{i,j}=t$, otherwise $t_{i,j}=rt$.
The sketch of the hopping $t_{i,j}$ is shown in Fig.~\ref{fig:model}.
In the non-interacting case, the ground state is a quantum spin Hall insulator (QSHI) with spin Chern number two for finite $t_{so}$. 
Accordingly, we can observe the helical edge modes localized around $i_x=0$ and $i_x=L-1$ under the open (periodic) boundary condition for $x$- ($y$-) direction, respectively.
These helical edge modes are protected by the time-reversal, charge $\mbox{U}(1)$, and spin $\mbox{U}(1)$ symmetry.
%%%%%%%%%%%%%%%%%%%%%%%%%
%%%%%%%%%%%%%%%%%%%%%%%%%

Our interest in this paper is how the correlations affect the two copies of QSHIs under the cylinder geometry. 
Real-space dynamical mean field theory, an extended version of the dynamical mean field theory\cite{DMFT_Muller_1989,DMFT_Metzner_PRL1989,DMFT_Georges_1996}, provides insights into inhomogeneous correlated systems\cite{RDMFT_Potthoff,RDMFT_Okamoto,RDMFT_Snoek}. 
A sketch of this approach is shown in Fig.~\ref{fig:model}.
Following this framework, we map the system to $L$-effective impurity models and compute the self-energy $\Sigma_{i_x\alpha\sigma}$ self-consistently. Here $\Sigma_{i_x\alpha\sigma}$ is the self-energy with spin $\sigma=\uparrow,\downarrow$ at site $i$ in layer $\alpha=a,b$.
The self-consistent equation is written as
%%%%%%%%%%%%%%%%%%
\begin{subequations}
\begin{eqnarray}
\hat{\mathcal{G}}^{-1}_{\alpha\sigma}(i\omega_n) &=&[\sum_{k_y}\hat{G}(k_y,i\omega)]^{-1}+\hat{\Sigma}_{\alpha \sigma}(i\omega_n), \\
\hat{G}(k_y,i\omega)&=&\{i\omega_n \1 -\hat{h}_{\alpha\sigma}(k_y)  -\hat{\Sigma}_{\alpha \sigma}(i\omega_n)\}^{-1}, 
\end{eqnarray}
with diagonal matrices
\begin{eqnarray}
\!\!\!\hat{\Sigma}_{\alpha \sigma}(i\omega_n)&:=&\mathrm{diag}\left(\Sigma_{0\alpha\sigma},\cdots, \Sigma_{i\alpha\sigma} ,\cdots,\Sigma_{L-1\alpha\sigma}\right), \\
\!\!\!\hat{\mathcal{G}}_{\alpha\sigma}(i\omega_n)&:=&\mathrm{diag}\left(\mathcal{G}_{0\alpha\sigma}, \cdots, \mathcal{G}_{i\alpha\sigma}, \cdots, \mathcal{G}_{L-1\alpha\sigma}\right),
\end{eqnarray}
where $\hat{h}_{\alpha\sigma}(k_y)$ is the Fourier transform of the hopping matrix.
\end{subequations}
%%%%%%%%%%%%%%%%%%
The self-energy $\Sigma_{i\alpha\sigma}$ is calculated in the effective impurity model at site $i_x$ from given effective Green's function $\mathcal{G}_{i_x\alpha\sigma}(i\omega_n)$. 
The partition function of the effective impurity model at site $i_x$ is given by

%%%%%%%%%%%%%%%%%%
\begin{subequations}
\begin{eqnarray}
\mathcal{Z}_{\mathrm{imp},i_x}&=&\int \Pi_{\sigma} \mathcal{D}\hat{\Psi}_{i_x,\sigma}\mathcal{D}\hat{\bar{\Psi}}_{i_x,\sigma} e^{-S_\mathrm{imp}}, \\
\mathcal{S}_{\mathrm{imp},i_x}&=&-\int d\tau d\tau' [ \sum_{\sigma,\alpha} \bar{c}_{i_x\alpha \sigma}(\tau) \mathcal{G}^{-1}_{i_x,\alpha}(\tau-\tau') c_{i_x\alpha \sigma}(\tau') \nonumber\\
&& \quad\quad\quad\quad\quad  + H_{\mathrm{imp},i_x} ], \\
H_{\mathrm{imp},i_x} &=& U\sum_{\alpha} (n_{i_x\alpha\uparrow}-\frac{1}{2}) (n_{i_x\alpha\downarrow}-\frac{1}{2}) +J\bm{S}_{i_xa}\cdot \bm{S}_{i_xb}, \nonumber \\
\end{eqnarray}
\end{subequations}
%%%%%%%%%%%%%%%%%%
with the path integral formalism.
Here, $\hat{\Psi}_{i_x,\sigma}:=(c_{i_xa\sigma},c_{i_xb\sigma})$ is the Grassmannian fermion field in layer $a$ and $b$ with spin $\sigma=\uparrow,\downarrow$,  and  $\tau\in [0,\beta]$ denotes the imaginary time.
This impurity model is essentially the same as a two-orbital impurity model.
In order to compute the self-energy, $\Sigma_{i_x\alpha\sigma}$, we employ the continuous-time quantum Monte Carlo (CTQMC) which is a powerful tool to analyze multi-orbital systems\cite{Werner_CTQMC_PRL2006,Werner_CTQMC_CTQMC2006,Haule_CTQMC_CTQMC2007}.

In order to characterize the topological properties, we calculate the spin-Hall conductivity with the Kubo formula:
%%%%%%%%%%%%%%%%%%
\begin{subequations}
\begin{eqnarray}
\sigma^{xy}_\mathrm{spin}&=& \frac{e}{2\pi} N_\mathrm{sCh}, \\
N_\mathrm{sCh} &=& \sum_{\sigma}\int d^3k \frac{\mathrm{sgn}(\sigma)\epsilon^{\mu\nu\rho}}{2(2\pi)^3} \mathrm{tr}[G^{-1}_\sigma\partial_{\mu}G_\sigma \nonumber \\
&&\quad \quad \quad \quad \quad \quad  \times G^{-1}_\sigma\partial_{\nu} G_\sigma G^{-1}_\sigma \partial_{\rho} G_\sigma],
\end{eqnarray}
\end{subequations}
%%%%%%%%%%%%%%%%%%
where $\hat{G}_{\sigma}:=\hat{G}_{\sigma}(i\omega,\bm{k})$ is the $2\times 2$ Green's function for the bulk, 
$\epsilon^{\mu\nu\rho}$ is the anti-symmetric tensor ($\mu,\nu,\rho=0,1,2$, $\epsilon^{012}=1$), $\bm{\partial}:=(\partial_{\omega},\partial_{k_x},\partial_{k_y})$, and $\bm{k}:=(k_x,k_y)$. 
$\mathrm{sgn}(\sigma)$ takes $1$ ($-1$) for $\sigma=\uparrow$ ($\downarrow$), respectively.
We have used the convention of summing repeated indices.
 At $T=0$, the spin-Hall conductivity is proportional to the spin Chern number (integer) $N_\mathrm{sCh}$, which characterizes the topological properties of the map from $(\omega,\bm{k})$ to $GL(2,\mathbb{C})$, even in the presence of electron correlations\cite{Ishikawa_IQHE_1987,Haldane_hall2004}. The spin Chern number is well-defined when the Green's function is non-singular, [i.e., $\mathrm{det}\hat{G}_{\sigma}(i\omega,\bm{k})\neq0$ and $\mathrm{det}\hat{G}^{-1}_{\sigma}(i\omega,\bm{k})\neq0$ hold for arbitrary $(i\omega,\bm{k})$] \cite{Volovik_textbook2009,Gurarie_2011}. 
Note that the spin-Hall conductivity is not completely quantized at finite temperatures, but still gives a hallmark of the topological properties; if  $N_\mathrm{sCh}$ is close to a nonzero-integer (zero), we can say that the system is in a topological phase (trivial phase)\cite{TBI_Mott_Yoshida}. We will use this aspect of the spin Hall conductivity in the following discussions. In this paper, we set typical values of the parameters as follows: $T=0.05t$, $r=0.7$, $t_{so}=0.2t$, and $J=t$, which are adequate to discuss the emergence of the TEMI state.  Since we analyze the two-dimensional system at finite temperatures, no continuous symmetry is broken.

%%%%%%%%%%%%%%%%%%%%%%%%%
\section{Results}\label{sec: results}
%%%%%%%%%%%%%%%%%%%%%%%%%

\begin{figure}[!h]
\begin{center}
\includegraphics[width=\hsize,clip]{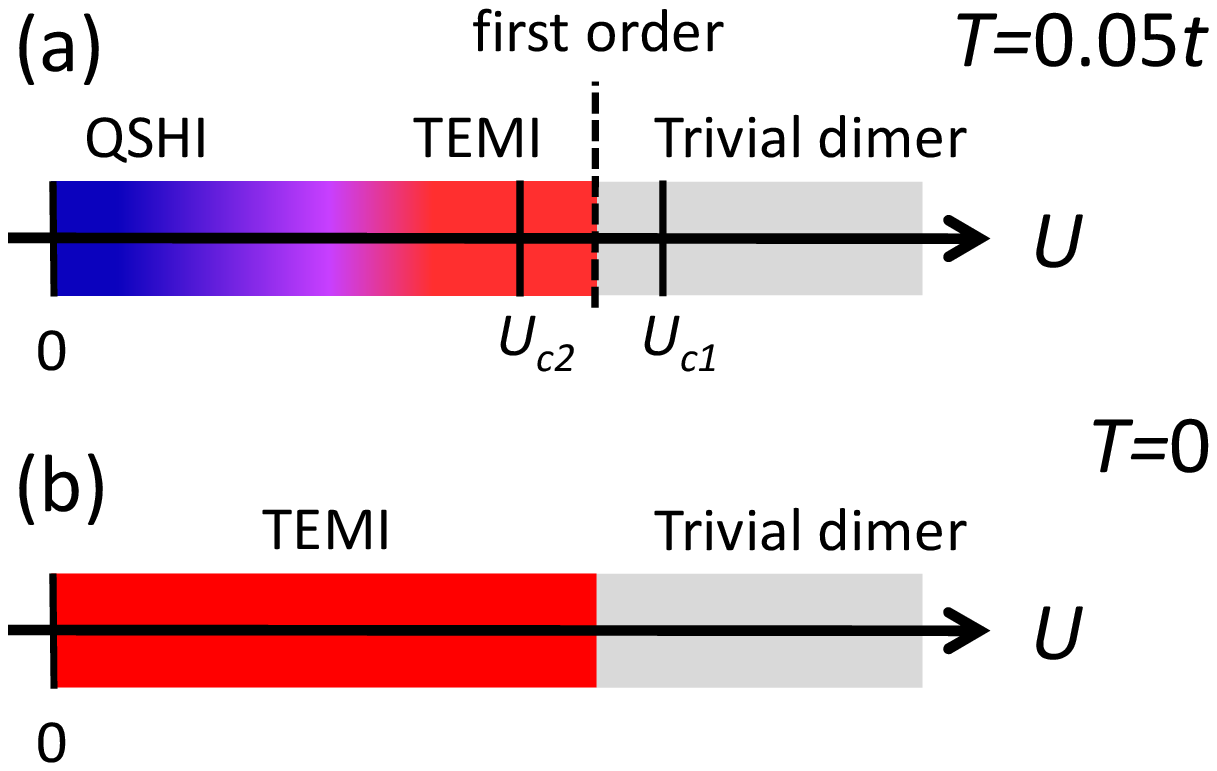}
\end{center}
\caption{
(Color Online). 
Phase diagram for $J=t$: (a) finite temperature ($T=0.05t$) and (b) zero temperature.  In (a), the quantum spin-Hall insulator (QSHI) is realized in the small $U$ region (blue region), which is gradually changed to the TEMI (read) via a crossover with increasing $U$. The TEMI state changes to a trivial dimer Mott state (the region colored with gray) via a first order transition at $U=U_c$. Correspondingly, the hysteresis behavior is observed for $U_{c2}<U<U_{c1}$, where $U_{c1}=4.25t$ and $U_{c2}=3.375t$.
In (b), one can see that the introduction of $J$ realizes a topological edge Mott insulating state (TEMI) at $T=0$, which is further driven to a trivial dimer Mott insulator with increasing $U$. These results at $T=0$ are in accordance with the analysis of (a) at finite temperatures.
}
\label{fig: phase}
\end{figure}

First of all, we briefly summarize our results. 
A typical phase diagram obtained at finite temperatures is shown in Fig.~\ref{fig: phase}(a), which is derived from the R-DMFT analysis with CTQMC at $T=0.05t$ (see Sec.~\ref{sec: numerical_results}). In this phase diagram, we use the terms, QSHI and TEMI, for topological phases, which are strictly defined only at $T=0$. Even at finite temperatures, we can still see clear remnants of topological properties. Our definition of the QSHI and the TEMI at finite temperatures is as follows. These two states are both characterized by the almost quantized spin-Hall conductivity $\sigma^{xy}_\mathrm{spin}\sim2(e/2\pi)$. The difference between them is that the edge modes of the QSHI are gapless (helical edge modes) while those of the TEMI are gapful in the single-particle spectrum. In the latter case, only the collective spin excitations become gapless, which is the definition of the TEMI. These two states are smoothly connected with increasing $U$ at finite temperatures via a crossover where the single-particle spectrum gradually acquires a gap for larger $U$.  Further increase in $U$ eventually induces a first order phase transition at $U=U_c$ from the TEMI to a trivial dimer Mott phase, where electrons in layer $a$ and $b$ form a singlet at each site and thus the topological properties are completely destroyed 
(With lowering temperature, the order of the phase transition can change to second order due to strong spatial fluctuation\cite{bilayerKane_Slagle_2015,bilayerKane_He_2015}). This first order transition accompanies the hysteresis behavior and the coexistence phase for $U_{c2}<U<U_{c1}$.
These R-DMFT analyses clearly demonstrate the emergence of the TEMI state.

We note that lowering temperature shifts the crossover region in Fig.~\ref{fig: phase}(a) to smaller $U$. So, what happens at zero temperature? To address this question, we show the phase diagram at $T=0$ in Fig.~\ref{fig: phase}(b), which is obtained with a bosonization approach for edge modes in one dimension (see Sec.~\ref{sec: edge}).  The bosonization analysis indicates that the TEMI state emerges for any $J>0$, because at $T=0$ the spin exchange interaction $J$ is marginally relevant for edge states and induces a gap which monotonically increases with increasing $J$. The nontrivial properties of the TEMI state induce a collective helical mode in $\langle n_{\sigma}(x) n _{\sigma}(0)\rangle$ with $n_{\sigma}:=\sum_{\alpha}n_{\alpha\sigma}$ ($\sigma=\uparrow,\downarrow$) in accordance with the above finite-temperature results. The TEMI phase at $T=0$ persists up to the phase transition to the trivial dimer Mott phase in the large $U$ region, as already mentioned in Fig.~\ref{fig: phase}(a). 

Below, we describe how we have arrived at these conclusions.
In Sec.~\ref{sec: numerical_results}, we first discuss details of the phase diagram [Fig.~\ref{fig: phase}(a)] obtained by R-DMFT with CTQMC at finite temperatures, and then in Sec.~\ref{sec: edge} address the zero temperature properties [Fig.~\ref{fig: phase}(b)].

%%%%%%%%%%%%%%%%%%%%%%%%%
%%%%%%%%%%%%%%%%%%%%%%%%%

%%%%%%%%%%%%%%%%%%%%%%%%%
\subsection{R-DMFT analysis at finite temperatures}\label{sec: numerical_results}
%%%%%%%%%%%%%%%%%%%%%%%%%

%%%%%%%%%%%%%%%%%%%%%%%%%%%%%%%%%%%%%%%%%%%%%%
\subsubsection{ Topological edge Mott insulator and crossover behavior}
%%%%%%%%%%%%%%%%%%%%%%%%%%%%%%%%%%%%%%%%%%%%%%

\begin{figure}[!h]
\begin{minipage}{1.0\hsize}
\begin{center}
\includegraphics[width=\hsize,clip]{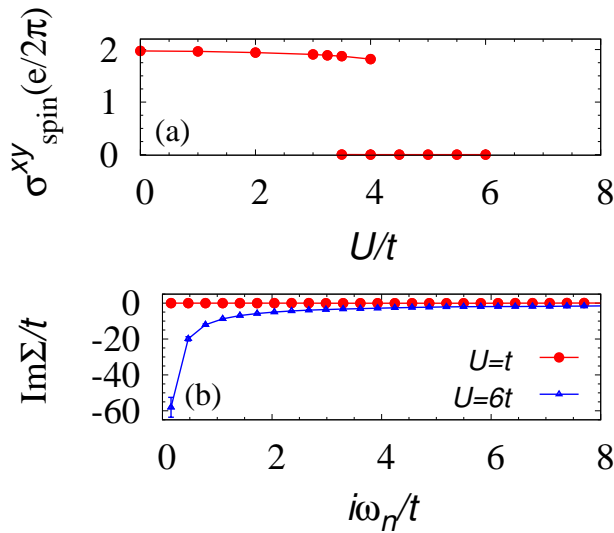}
\end{center}
\end{minipage}
\caption{
(Color Online). 
(a) the spin-Hall conductivity as a function of interaction strength.
(b) imaginary part of the self-energy at $i_x=20$ for $U=t$ and $U=6t$.
In panel (a), we can find a hysteresis behavior, which indicates the presence of the first order transition.
}
\label{fig: cond}
\end{figure}

We start with bulk properties, i.e. properties at center of the cylinder ($i_x=20$).
In Fig.~\ref{fig: cond}(a), the spin-Hall conductivity at $T=0.05t$ is plotted as a function of the in-plane Hubbard interaction $U$ for a given choice of the inter-plane exchange interaction $J$ (we recall that the conductivity is proportional to the spin Chern number even in the strongly correlated systems at $T=0$\cite{Ishikawa_IQHE_1987,Haldane_hall2004}). We can see that the spin Hall conductivity is almost quantized, i.e. $\sigma^{xy}_\mathrm{spin}\sim2(e/2\pi)$ up to a critical value of $U=U_c$, beyond which it suddenly drops to zero via a first order transition. Therefore we can say that the system is in a topologically nontrivial state for $U<U_c$, while it is changed to a trivial phase for $U>U_c$. However, only from the data for $\sigma^{xy}_\mathrm{spin}$, we cannot figure out which kind of topological phase the system belongs to for $U<U_c$. This will be clarified through the analysis of the edge states below.
%%%%%%%%%%%%%%%%%%%%%%%%%
%%%%%%%%%%%%%%%%%%%%%%%%%

Here we make a brief comment on how the topological properties disappear at the first order transition point in the bulk. This phase transition is due to zeros of the Green's function (or the divergence of the self-energy), which completely destroy the topologically nontrivial structure. In Fig.~\ref{fig: cond}(b), we plot the imaginary part of the self-energy at $i_x=20$ for $U=t$ and $U=6t$. 
For $U=6t$, the self-energy indeed diverges (i.e., zero of Green's function appears), while for $U=t$ the self-energy is non-singular and thus the system behaves as a renormalized band insulator, in accordance with the almost quantized spin-Hall conductivity for $U<U_c$ shown in Fig. \ref{fig: cond}(a).
%%%%%%%%%%%%%%%%%%%%%%%%%
\begin{figure}[!h]
\begin{minipage}{0.75\hsize}
\begin{center}
\includegraphics[width=\hsize,clip]{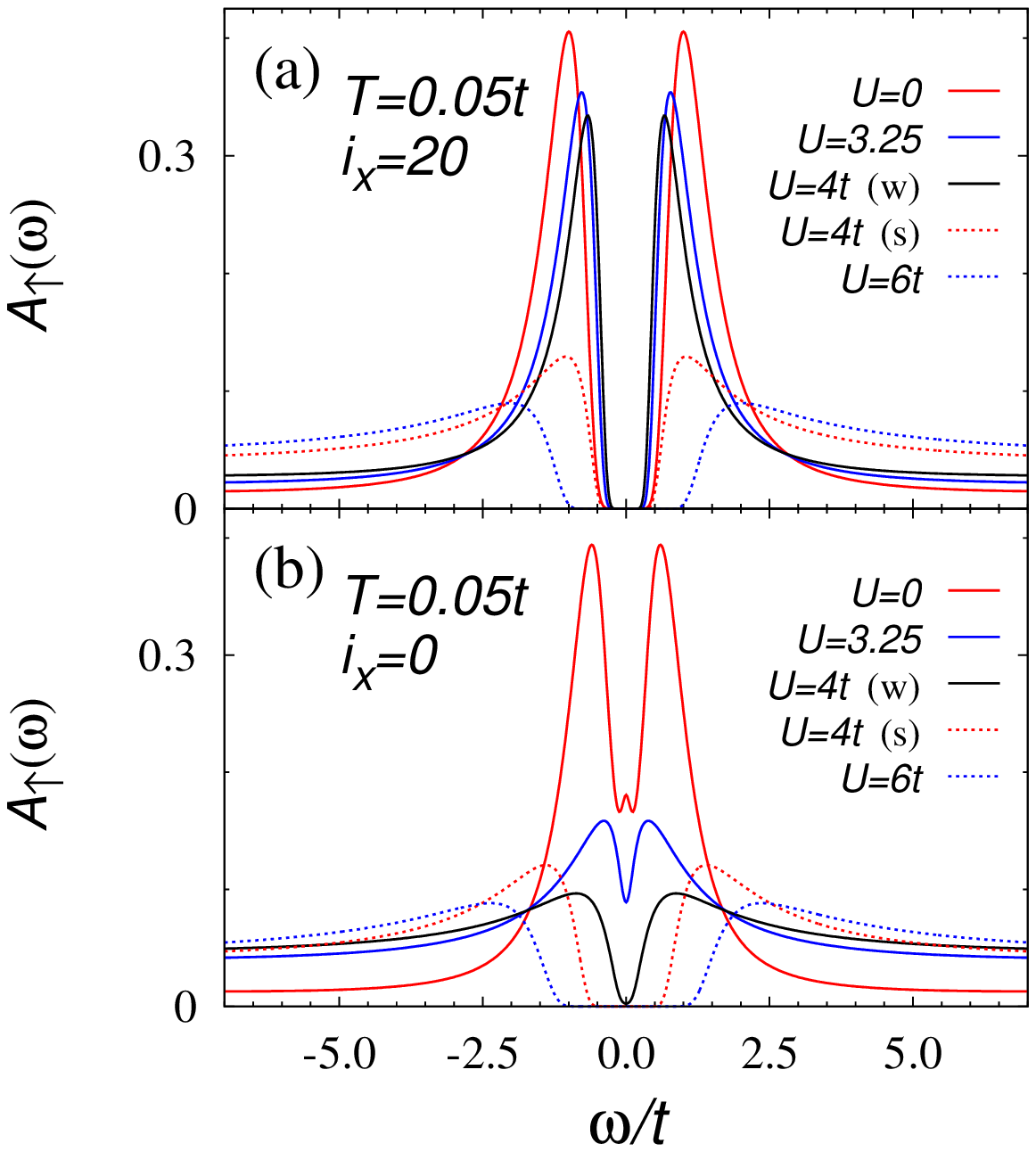}
\end{center}
\end{minipage}
\begin{minipage}{1.0\hsize}
\begin{center}
\includegraphics[width=\hsize,clip]{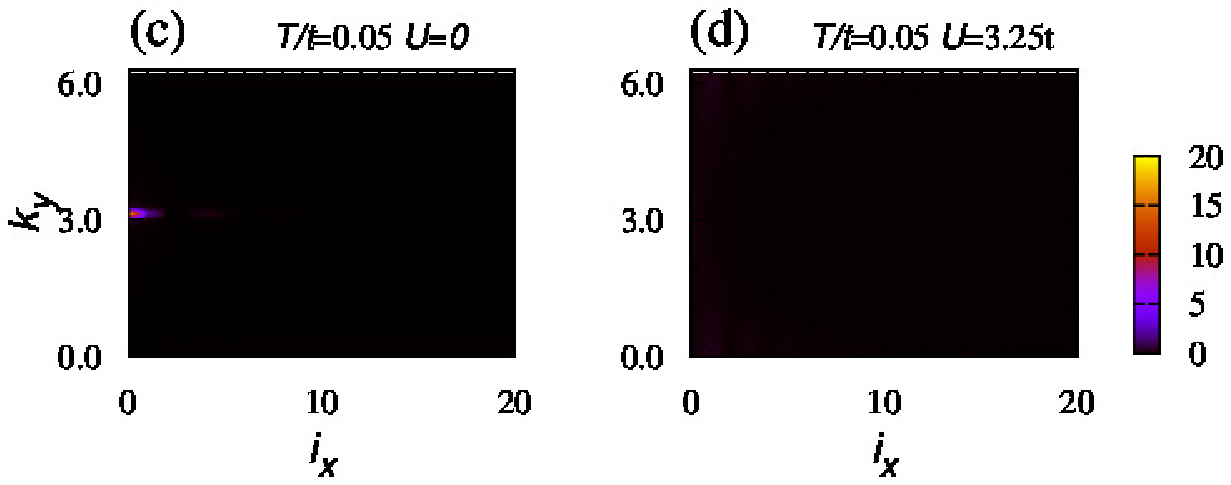}
\end{center}
\end{minipage}
\caption{
(Color Online). 
(a) and (b): local density of states (LDOS) at $i_x=0$ and $i_x=20$ for several values of interaction strength.
For $U=4t$, we can obtain two solutions associated with hysteresis behavior.
Data labeled with ``(w)" are obtained by inputting a solution of weak coupling region for the R-DMFT loop, while the ones labeled with ``(s)" are obtained with a solution of strong coupling region. 
The former (latter) describes the QSHI phase (trivial dimer Mott phase), respectively. 
(c) and (d): momentum-resolved spectral weight $\left( -\mathrm{Im} [\hat{G}(k_y,\omega=0)]_{i_x,i_x}/\pi \right)$ at the Fermi energy for $U=0$ and $J=3.25t$.
In panel (c), we can find a point like peak around $(i_x,k_y)=(0,\pi)$ which means the presence of helical edge modes in the single particle spectrum. On the other hand no peak is observed in panel (d).
In these panels, the spectral weight for $0\leq i_x \leq 20$ is plotted because the data are symmetric with respect $i_x=20$.
}
\label{fig: DOS_and_Ak}
\end{figure}
%%%%%%%%%%%%%%%%%%%%%%%%%

Let us now turn to the edge properties. First recall that in the non-interacting case ($U=J=0$), the quantized conductivity $\sigma^{xy}_\mathrm{spin}\sim2(e/2\pi)$ predicts two pairs of gapless helical edge modes as long as the symmetry is respected. Since the present results obtained at finite temperatures have no spontaneous symmetry breaking, one might expect that the gapless helical edge modes exist in the single-particle spectrum in the region where $\sigma^{xy}_\mathrm{spin}\sim 2(e/2\pi)$ holds. 
Remarkably, however, this is not the case in the presence of the spin exchange interaction $J$ and the Hubbard interaction $U$.  To clarify
 this point, in Fig.~\ref{fig: DOS_and_Ak}, we plot the local density of states (LDOS) $A_\uparrow(\omega)$ and the momentum-resolved spectral weight at the Fermi energy $A_{\uparrow}(k_y,i_x,\omega=0)$. 
While the LDOS at a bulk site shown in Fig.~\ref{fig: DOS_and_Ak}(a) indicates the presence of bulk gap for all the parameters chosen, the LDOS at edge sites ($i_x=0$) plotted in Fig.~\ref{fig: DOS_and_Ak}(b) strongly depends on the values of $U$. 
For $U=0$, we find a peak at the Fermi energy $\omega=0$, indicating that there are gapless edge modes. 
With increasing $U$, however, this peak gradually changes to a gap structure. For example, as seen in Fig.~\ref{fig: DOS_and_Ak}(b), the peak changes to a pseudo-gap for $U=3.25t$ and to a gap for $U=4t$, implying that the helical edge modes in the single-particle channel are destroyed gradually. Such destruction of the edge modes in the single-particle spectrum is more clearly seen in the momentum resolved spectral weight $A_{\uparrow}(k_y,i_x,\omega=0)$ plotted in Figs.~\ref{fig: DOS_and_Ak}(c) and (d). 
For $U=0$, the system shows helical edge modes in the single-particle spectrum, resulting in a sharp peak of the momentum resolved spectral weight at $k_y=\pi$
 [see Fig.~\ref{fig: DOS_and_Ak}(c)], which is consistent with the presence of the peak at $\omega=0$ in the LDOS [Fig.~\ref{fig: DOS_and_Ak}(b)].
On the other hand, the gapless modes are destroyed for $U=3.25t$, where
the spectral weight $A_{\uparrow}(k_y,i_x=0,\omega=0)$ does not show any peak at $k_y=\pi$, indicating that the helical edge modes in the single particle-spectrum are gapped out without symmetry breaking.

Putting the above bulk and edge properties together, we conclude that the crossover behavior appears in two types of topological phases at a given temperature. In smaller $U$, we find the QSHI, which is characterized by the almost quantized  spin-Hall conductivity $\sigma^{xy}_\mathrm{spin}\sim 2(e/2\pi)$ and the gapless helical edge modes. On the other hand, for larger $U$, the spin-Hall conductivity is still almost quantized, but the single-particle spectrum does not host any gapless mode, indicating the presence of gapless modes carrying the spin current only in collective excitation spectra.  This phase is thus identified as the TEMI.

The above crossover between the QSHI (smaller $U$) and the TEMI (larger $U$) is observed at a given temperature ($T=0.05t$), which is much less than the size of bulk energy gap. When the temperature increases, the crossover becomes obscured, and at temperatures higher than  the bulk energy gap, the system enters the high temperature region where all the topological properties disappear.  On the other hand, when the temperature becomes lower, we can confirm that the crossover region shifts to the small $U$ region; namely the QSHI (TEMI) region becomes smaller (larger). This is because at lower temperatures, the effects of $J$ and $U$ are more prominent, leading to the enhancement of the gap at edge modes and thus stabilizing the TEMI. 
So, we naturally expect that the TEMI dominates the topological phase.
In Sec.~\ref{sec: edge}, we will address this problem and find that topological phase at $T=0$ is indeed in the TEMI state as far as $J$ is finite.

%%%%%%%%%%%%%%%%%%%%%%%%%%%%%%%%%%%%%%%%%%%%%%
\subsubsection{  Phase competition}
%%%%%%%%%%%%%%%%%%%%%%%%%%%%%%%%%%%%%%%%%%%%%%
 
Here, we discuss the phase competition between the topologically nontrivial phase (TEMI) and the trivial  Mott phase. As mentioned above, in the presence of the spin exchange interaction $J$, increasing the Hubbard interaction $U$ firstly induces 
a crossover from the QSHI to the TEMI and then drives a first order transition to the trivial dimer Mott insulator where the single-particle gap is of the order of $U$ at every site.
%%%%%%%%%%%%%%%%%%%%%%%%%
\begin{figure}[!h]
\begin{minipage}{1.0\hsize}
\begin{center}
\includegraphics[width=\hsize,clip]{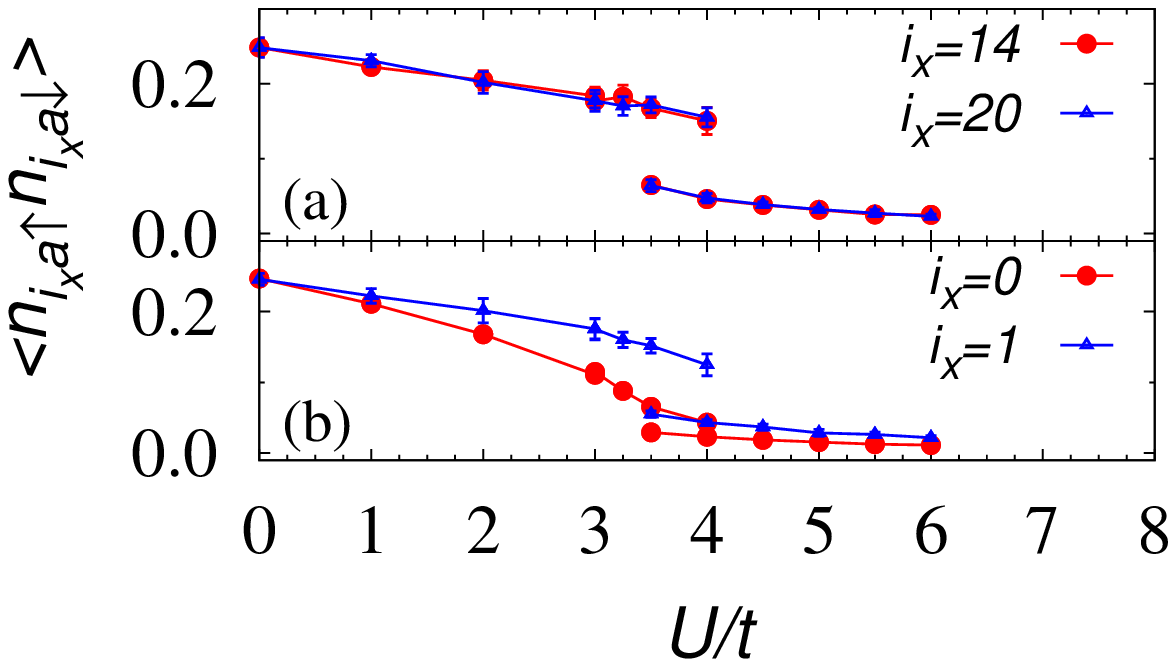}
\end{center}
\end{minipage}
\begin{minipage}{1.0\hsize}
\begin{center}
\includegraphics[width=\hsize,clip]{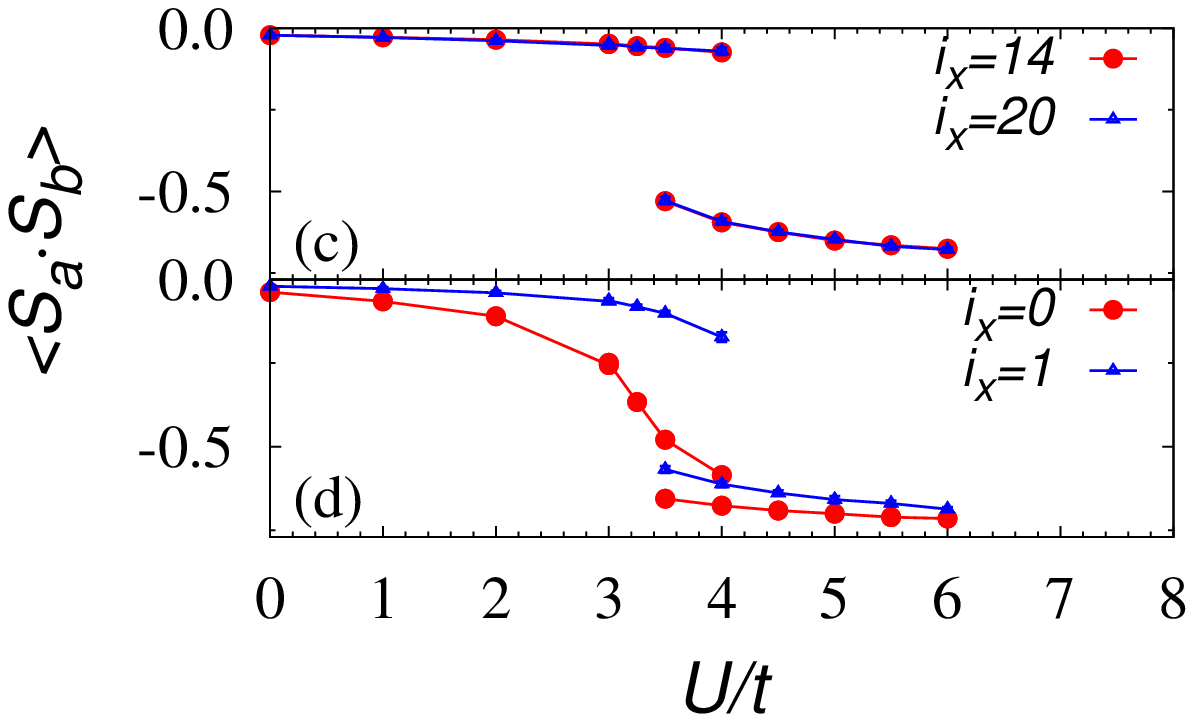}
\end{center}
\end{minipage}
\caption{(Color Online). 
(a) [(b)]: Interaction dependence of the double occupancy at $i_x=14$ and $20$ ($i_x=0$ and $1$), respectively. 
(c) [(d)]: Interaction dependence of the inter-layer local spin correlation at $i_x=14$ and $20$ ($i_x=0$ and $1$), respectively.
}
\label{fig: interaction_Docc}
\end{figure}
%%%%%%%%%%%%%%%%%%%%%%%%%

 We start with the first order transition. 
The double occupancy and the local spin correlation at site $i_x$ in layer $a$ are plotted in Fig.~\ref{fig: interaction_Docc}.
In the region of small $U$, the double occupancy and the spin correlation gradually decrease at every site with increasing the interaction strength, in accordance with the crossover behavior between the QSHI and the TEMI. 
Further increase in $U$ drives  the system  to a trivial phase from the TEMI via the first order transition. 
 The topologically trivial phase for $U>U_c$ is a dimer Mott phase for which electrons in layer $a$ and $b$ form a singlet at each site.
The first order transition is accompanied by the hysteresis in the interaction dependences of the double occupancy and the local spin correlation for $U_{c2}<U< U_{c1}$, where $U_{c1}=4.25t$ and $U_{c2}\sim 3.375t$. 
Figure ~\ref{fig: interaction_Docc} shows jumps  in the double occupancy at $U_{c1}$ and $U_{c2}$ at every site (the jumps are more clearly observed in the interaction dependence of the spin correlation).
%%%%%%%%%%%%%%%%%%%%%%%%%
\begin{figure}[!h]
\begin{minipage}{1.0\hsize}
\begin{center}
\includegraphics[width=\hsize,clip]{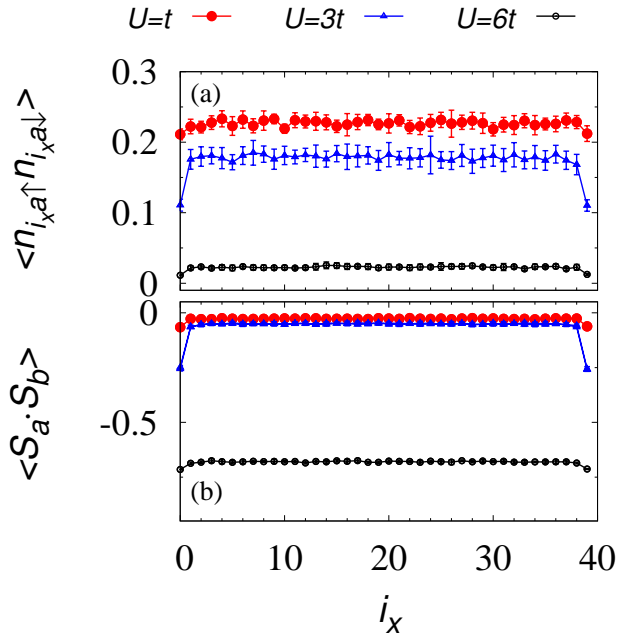}
\end{center}
\end{minipage}
\caption{(Color Online). 
Spatial dependence of (a) double occupancy and  (b) spin correlation.
}
\label{fig: sites_Docc}
\end{figure}
%%%%%%%%%%%%%%%%%%%%%%%%%
We note that the two adjacent insulating states separated with the first order phase transition at $U_c$ have totally different properties from each other although both of them have the single-particle gap in the edge modes.
For the small $U$ side (topological phase), the Mott behavior is observed only at edge sites, while  for the large $U$ side (trivial phase), the Mott behavior is observed at both of bulk and edge sites. In the former topological region, we have already observed that the helical edge modes are destroyed while the bulk behaves as a renormalized band insulator [see Figs.~\ref{fig: DOS_and_Ak}(a) and (b)]. 
Correspondingly, in this topological region, only electrons around the edges are strongly affected, as clearly seen in Figs.~\ref{fig: sites_Docc} where spatial dependences of the double occupancy and spin correlation are plotted. For $U=3t$ in the topological phase, the electrons around edges show the dramatic suppression of the double occupancy  [Figs.~\ref{fig: sites_Docc} (a)] and also the strong singlet correlation  [Figs.~\ref{fig: sites_Docc} (b)] while the other electrons in the bulk are less correlated.  
On the other hand, in the large $U$ trivial region, whole the system shows a trivial Mott insulating behavior. Actually, as seen from the data for $U=6t$, the double occupancy has small values and the singlet correlation is strong at every site. Correspondingly, both of the single-particle gap in the bulk and at the edges are of the order of $U$ (i.e. the Mott gap)  [see Figs.~\ref{fig: DOS_and_Ak}(a) and (b)]. 

Summarizing all the above results, we end up with  the phase diagram shown in Fig.~\ref{fig: phase}(a), which supports the crossover between the QSHI state and the TEMI state, and also the phase transition to the trivial Mott insulator at finite temperatures.

%%%%%%%%%%%%%%%%%%%%%%%%%%%%%%%%%%%%%%%%%%%%%%
\subsection{
Formation of topological edge Mott insulator at zero temperature
}\label{sec: edge}
%%%%%%%%%%%%%%%%%%%%%%%%%%%%%%%%%%%%%%%%%%%%%%

So far we have been concerned with the finite temperature properties. 
At zero temperature, we expect that infinitesimal interaction $J (>0)$ induces the TEMI state, which persists even in the presence of $U$ as far as the system is in a topological phase. This is supported here by the analysis based on the bosonization approach. To this end, we start with the model (\ref{fig:model}) with $J=U=0$, and discuss how the instability of gapless edge modes occur in the presence of $J$ and $U$. We use the Chern-Simons theory and the bosonization method\cite{Lu_CS_2011,Isobe_Fu2015,You_bosonization_2015}, following 
the approach used in Ref.~\onlinecite{You_bosonization_2015}.

Consider the edge states of two copies of non-interacting QSHI with spin Chern number one. Then its effective action is given by
%%%%%%%%%%%%%%%%%%
\begin{subequations}
\label{eq: S_0}
\begin{eqnarray}
S_{edge}&=& \int \frac{dx d\tau}{4\pi}    \{iK_{I,J} \partial_{\tau}\bm{\phi}^I(t,x)\partial_{x}\bm{\phi}^J(t,x) \nonumber \\
        && \quad\quad +V_{I,J}\partial_{x}\bm{\phi}^I(t,x) \partial_{x}\bm{\phi}^J(t,x)\}, 
\end{eqnarray}
with
\begin{eqnarray}
c_{\alpha\sigma}(x)=\frac{\kappa_{\alpha\sigma}}{\sqrt{2\pi\alpha_0}}e^{i\mathrm{sgn}(\sigma)k_Fx}e^{-i\phi_{\alpha\sigma}(x)},\\
n_{\alpha\sigma}(x)=-\frac{\mathrm{sgn}(\sigma)}{2\pi} \partial_x \phi_{\alpha\sigma}(x),
\end{eqnarray}
\end{subequations}
%%%%%%%%%%%%%%%%%%
where, $\bm{\phi}^T:=(\phi_{a\uparrow},\phi_{a\downarrow},\phi_{b\uparrow},\phi_{b\downarrow})$ are bosonic fields.
$K=\sigma^z\oplus\sigma^z$, and $V=v_F\sigma^z\oplus\sigma^z$. Here $k_F$ and $v_F$ are the Fermi momentum and the Fermi velocity of edge states, respectively.
$\kappa_{\alpha\sigma}$ denotes the Klein factor with $\alpha=a,b$ and $\sigma=\uparrow,\downarrow$, and $\alpha_0$ is a positive constant.
$\mathrm{sgn}(\sigma)$ takes 1 ($-1$) respectively for $\sigma=\uparrow$ ($\downarrow$). 
We note that $\phi_1$ and $\phi_3$ ($\phi_2$ and $\phi_4$) describe the right (left) movers.

Under the symmetry operations, these fields are transformed as
%%%%%%%%%%%%%%%%%%
\begin{eqnarray}
T
\bm{\phi}
T^{-1}
&=&
-\sigma^x\oplus\sigma^x
\bm{\phi}
+
\left(
\begin{array}{c}
0   \\
\pi \\
0   \\
\pi
\end{array}
\right),
\nonumber \\
u_{c}(\theta_c) 
\bm{\phi}
u^{-1}_{c}(\theta_c) 
&=&
\bm{\phi}
+
\theta_c
\left(
\begin{array}{c}
1  \\
1  \\
1  \\
1
\end{array}
\right),
\nonumber \\
u_{s}(\theta_s) 
\bm{\phi}
u^{-1}_{s}(\theta_s) 
&=&
\bm{\phi}
+
\theta_s
\left(
\begin{array}{c}
1  \\
-1 \\
1 \\
-1 \
\end{array}
\right),
\end{eqnarray}
%%%%%%%%%%%%%%%%%%
where $T$, $u_c(\theta_c)$, and $u_s(\theta_s)$ denote the operators for the time-reversal, charge U(1) rotation, and spin U(1) rotation, respectively.

In terms of these bosonic fields, the interaction terms can be written as follows:
%%%%%%%%%%%%%%%%%%
\begin{eqnarray}
U\sum_{\alpha}n_{\alpha\uparrow}n_{\alpha\downarrow} &=&-\frac{U}{(2\pi)^2}\sum_{\alpha}\partial_x\phi_{\alpha\uparrow}\partial_x\phi_{\alpha\downarrow}, \\
JS^z_aS^z_b &=& \frac{J}{4(2\pi)^2}\sum_{\sigma,\sigma'}(\partial_x \phi_{a\sigma})(\partial_x \phi_{b\sigma'}), \\
\frac{J}{2}(S^+_aS^-_b+h.c.) &=& \frac{J\cos(\phi_{a\uparrow}-\phi_{a\downarrow}-\phi_{b\uparrow}+\phi_{b\downarrow})}{(2\pi\alpha_0)^2},
\end{eqnarray}
%%%%%%%%%%%%%%%%%%
Thus, the effective action in the presence of the interactions is written as
%%%%%%%%%%%%%%%%%%
\begin{subequations}
\label{eq: S_tot}
\begin{eqnarray}
S_{edge}&=&  \int \frac{dx d\tau}{4\pi} \{iK_{I,J} \partial_{\tau}\bm{\phi}^I(t,x)\partial_{x}\bm{\phi}^J(t,x) \nonumber \\
        && \quad\quad +\left(V+V_\mathrm{int}\right)_{I,J}\partial_{x}\bm{\phi}^I(t,x) \partial_{x}\bm{\phi}^J(t,x) \} \nonumber \\
        && +\frac{J}{(2\pi\alpha_0)^2}\int dx d\tau \cos(\phi_{a\uparrow}-\phi_{a\downarrow}-\phi_{b\uparrow}+\phi_{b\downarrow}), \nonumber \\
\end{eqnarray}
with
\begin{eqnarray}
V_\mathrm{int}&=&
\frac{1}{2\pi}
\left(
\begin{array}{cccc}
0 & -U & J/4 & J/4 \\
-U & 0 & J/4& J/4 \\
J/4 & J/4& 0&-U  \\
J/4 & J/4 & -U&0 
\end{array}
\right).
\end{eqnarray}
\end{subequations}
%%%%%%%%%%%%%%%%%%
Since $[(\phi_1-\phi_2-\phi_3+\phi_4)(x),(\phi_1-\phi_2-\phi_3+\phi_4)(x')]=0$ holds, the last term in Eq.~(\ref{eq: S_tot}a) pins the field $\phi_1-\phi_2-\phi_3+\phi_4$ when this term becomes relevant.
We recall that this interaction term is symmetry allowed [i.e., this is invariant under applying $T$, $u_c(\theta_c)$ and $u_s(\theta_s)$] and pins the field without symmetry breaking.

Now, we apply the renormalization procedure.
As a first step, we change the basis as follows:
%%%%%%%%%%%%%%%%%%
\begin{subequations}
\begin{eqnarray}
\psi&=&M^{-1}\bm{\phi},
\end{eqnarray}
\end{subequations}
with 
\begin{eqnarray}
M&:=&\frac{1}{2}
\left(
\begin{array}{cccc}
 1 & 1 & 1 & 1  \\
 1 &-1 &-1 & 1  \\
 1 & 1 &-1 &-1  \\
 1 &-1 & 1 &-1
\end{array}
\right).
\end{eqnarray}
%%%%%%%%%%%%%%%%%%
As a result, the $K$-matrix, $V+V_\mathrm{int}$, and $\cos(\phi_1-\phi_2-\phi_3+\phi_4)$ are transformed as
%%%%%%%%%%%%%%%%%%
\begin{subequations}
\begin{eqnarray}
&&K\to \sigma^x\otimes\sigma^x, \\
&&V+V_\mathrm{int}\to \mathrm{diag}\left(\frac{(J/2-U)}{2\pi}+v_F,\frac{U}{2\pi}+v_F, \right. \nonumber \\
&& \quad\quad\quad\quad\quad\left. \frac{U}{2\pi}+v_F, -\frac{(J/2+U)}{2\pi}+v_F \right), \\
&&\cos(\phi_1-\phi_2-\phi_3+\phi_4) \to\cos(2\psi_3).
\end{eqnarray}
\end{subequations}
%%%%%%%%%%%%%%%%%%
In this basis the action (\ref{eq: S_tot}) is separated into two sectors spanned by $(\psi_1,\psi_2)$ and $(\psi_3,\psi_4)$.
Only the second sector includes non-liner terms. 
Redefining the fields as $(\psi_3,\psi_4):=(\sqrt{2}\psi'_3,\sqrt{2}\psi'_4)$ and integrating out the fields $\psi'_4$, we obtain the sine-Gordon action:
%%%%%%%%%%%%%%%%%%
\begin{subequations}
\label{eq: S_SG}
\begin{eqnarray}
S_{SG}&=& \frac{1}{2\pi g} \int dx d\tau \left\{ \frac{1}{v_0}(\partial_\tau\psi'_3)^2+v_0(\partial_x\psi'_3)^2 \right\} \nonumber \\
&& \quad \quad +\frac{J}{(2\pi\alpha_0)^2}\int dx d\tau \cos(2\sqrt{2}\psi'_3), 
\end{eqnarray}
with
\begin{eqnarray}
v_0&=&2\sqrt{(v_F+\frac{U}{2\pi})(v_F-\frac{(U+J/2)}{2\pi})}, \\
g&=&\sqrt{\frac{2\pi v_F-(J/2+U)}{2\pi v_F+U}}.
\end{eqnarray}
\end{subequations}
%%%%%%%%%%%%%%%%%%

Appling the renormalization-group scheme to the sine-Gordon model (\ref{eq: S_SG}), we obtain the renormalization flow:
%%%%%%%%%%%%%%%%%%
\begin{subequations}
\label{eq: RGflow}
\begin{eqnarray}
\frac{dJ(l)}{dl}&\sim&J(l)[2-2g(l)], \\
\frac{dg(l)}{dl}&\sim&-J^2(l).
\end{eqnarray}
\end{subequations}
%%%%%%%%%%%%%%%%%%
$(g,J)=(1,0)$ is a fixed point. Expanding Eqs.~(\ref{eq: RGflow}) around this fixed point ($g=1+g'/2$), we obtain
%%%%%%%%%%%%%%%%%%
\begin{subequations}
\begin{eqnarray}
\frac{dJ(l)}{dl}&\sim&-J(l)g'(l), \\
\frac{dg'(l)}{dl}&\sim& -J^2(l).
\label{eq: RGflow2}
\end{eqnarray}
\end{subequations}
%%%%%%%%%%%%%%%%%%
Thus, for $J>0$, the non-linear term is marginally relevant and pins the field $\psi_3$ since $g<1$ holds. 
This is also the case in the presence of the Hubbard interaction $U$. 
Therefore, we conclude that with the infinitesimal antiferromagnetic spin exchange interaction, the single-particle excitations become massive without breaking symmetry.

The remaining gapless edge modes live in the subspace of $\psi_1$ and $\psi_2$.
In order to find excitation spectra showing the edge modes, we have to diagonalize the $K$-matrix for the remaining sector.
This can be done with redefining fields
%%%%%%%%%%%%%%%%%%
\begin{eqnarray}
\psi'_1&=& \frac{1}{\sqrt{2}}(\psi_1+\psi_2), \nonumber \\
\psi'_2&=& \frac{1}{\sqrt{2}}(\psi_1-\psi_2).
\end{eqnarray}
%%%%%%%%%%%%%%%%%%
Thus, we can find that $\langle \psi'_1(x) \psi'_1(0) \rangle$ and $\langle \psi'_2(x) \psi'_2(0) \rangle$ show gapless edge modes. 
Correspondingly, $\langle n_{\sigma}(x) n_{\sigma}(0)\rangle$ with $\sigma=\uparrow,\downarrow$, where $n_\sigma(x):=\sum_{\alpha}n_{\alpha\sigma} (x)$, hosts gapless mode propagating in opposite direction because Eq.~(\ref{eq: S_0}c) holds.  These modes are protected by symmetry because any potential term pinning $\psi_1$ or $\psi_2$ breaks the symmetry.

We note that the above results are obtained on the assumption that the spin Chern number takes two.  This assumption holds for the small $U$ region untill the trivial dimer Mott phase emerges. We thus end up with Fig.~\ref{fig: phase}(b).

%%%%%%%%%%%%%%%%%%%%%%%%%
\section{SUMMARY}\label{sec: discussion}
%%%%%%%%%%%%%%%%%%%%%%%%%

In this paper, we have analyzed a bilayer Kane-Mele-Hubbard model with the inter-layer spin exchange interaction under cylinder geometry by employing R-DMFT with CTQMC. Our systematic analysis for the bulk and the edge at finite temperatures has demonstrated the possible realization of the TEMI state whose topological structure remains nontrivial up to the Mott transition in the bulk, and induces gapless edge modes only in collective modes. 

The numerical data for bulk properties have indicated that the spin-Hall conductivity takes $\sigma^{xy}_\mathrm{spin}\sim2(e/2\pi)$ in the insulating phase, which predicts gapless edge modes carrying the spin current.
Intriguingly, numerical data for edge properties have suggested the presence of single-particle gap at edges in contrast to the non-interacting QSHI. 
Putting these observations for the bulk and edges together, we have ended up with the TEMI where collective modes carry the spin current.
With bosonization analysis, we have confirmed that the crossover region between the QSHI and the TEMI shifts to the smaller $U$ region with lowering temperature, and eventually at zero temperature the topological phase is completely dominated by the TEMI  for any $U<U_c$ if the antiferromagnetic spin exchange interaction $J$ is introduced. We have also shown that collective helical modes in density-density correlation function $\langle n_{\sigma}(x) n _{\sigma}(0)\rangle$ ($\sigma=\uparrow,\downarrow$) are protected by charge U(1), spin U(1), the time-reversal symmetry, where $n_{\sigma}$ denotes the density of electrons with spin $\sigma=\uparrow,\downarrow$.

In order to study low-temperature properties numerically in more detail, one needs to analyze the system by properly taking into account the spatial fluctuations which are considered to be important in the low temperature region.
The detailed analysis of the TEMI state is an open problem to be addressed, and also an extension of the system to three dimensions is an interesting future work.

\section{ACKNOWLEDGMENTS}
This work is partly supported by a Grand-in-Aid
for Scientific Research on Innovative Areas (JSPS KAKENHI Grant No. 15H05855) and also JSPS KAKENHI (No.
25400366). The numerical calculations were performed on
supercomputer at the ISSP in the University of Tokyo, and the SR16000
at YITP in Kyoto University.

%%%%%%%%%%%%%%%%%%%%%%%%%%%%%%%%%%%%%%
%%%%%%%%%%%%%%%%%%%%%%%%%%%%%%%%%%%%%%

\end{document}